%% file: manuscript.tex
\documentclass{article}
\usepackage{arxiv}
\usepackage[utf8]{inputenc} 
\usepackage[T1]{fontenc}    
\usepackage{hyperref}       
\usepackage{url}            
\usepackage{booktabs}       
\usepackage{amsfonts}       
\usepackage{nicefrac}       
\usepackage{microtype}      
\usepackage{fancyhdr}       
\usepackage{graphicx}       
\usepackage{float} 
\usepackage[T1]{fontenc}
\usepackage{dcolumn}
\usepackage{bm}
\usepackage{siunitx}
\usepackage{mathptmx}
\usepackage{etoolbox}
\usepackage{amsmath}
\usepackage{amssymb}
\usepackage{tabularx}
\usepackage{soul}
\setstcolor{red}
\DeclareSIUnit\angstrom{\text {Å}}
\usepackage{multirow}
\usepackage[style=chem-acs]{biblatex}
\title{Bayesian Analysis Reveals the Key to Extracting Pair Potentials from Neutron Scattering Data}

\author{B. L. Shanks, H. W. Sullivan, M. P. Hoepfner \\
Department of Chemical Engineering, University of Utah \\
Salt Lake City, UT\\
\texttt{B. L. Shanks} {brennon.shanks@chemeng.utah.edu}
}

\begin{document}

\maketitle

\begin{abstract}

The inverse problem of statistical mechanics is an unsolved, century-old challenge to learn classical pair potentials directly from experimental scattering data. This problem was extensively investigated in the 20th century but was eventually eclipsed by standard methods of benchmarking pair potentials to macroscopic thermodynamic data. However, it is becoming increasingly clear that existing force field models fail to reliably reproduce fluid structures even in simple liquids, which can result in reduced transferability and substantial misrepresentations of thermophysical behavior and self-assembly. In this study, we revisited the structure inverse problem for a classical Mie fluid to determine to what extent experimental uncertainty in neutron scattering data influences the ability to recover classical pair potentials. Bayesian uncertainty quantification was used to show that structure factors with random noise smaller than 0.005 to $\sim30$ $\si{\angstrom}^{-1}$ are required to accurately recover pair potentials from neutron scattering. Notably, modern neutron instruments can achieve this precision to extract classical force models to within approximately $\pm$ 1.3 for the repulsive exponent, $\pm$ 0.068 $\si{\angstrom}^{-1}$ for atomic size, and 0.024 kcal/mol in the potential well-depth with 95\% confidence. Our results suggest the exciting possibility of improving molecular simulation accuracy through the incorporation of neutron scattering data, advancement in structural modeling, and extraction of model-independent measurements of local atomic forces in real fluids. 

\end{abstract}

\section{Introduction}

Reconstructing interatomic potentials from experimental scattering data is a historic inverse problem in statistical mechanics, motivated by the idea that complete knowledge of the effective interatomic potential with the atomic correlation functions allows for all thermodynamic properties of a classical liquid to be calculated \cite{kirkwood_statistical_1951}. While it has become widely accepted that liquid state systems exhibit significant many-body and quantum mechanical (both electronic and nuclear) interactions \cite{hansen_theory_2013} that influence molecular dynamics, the fact that empirical molecular simulations remain the gold-standard for efficient and accurate liquid state materials modeling has maintained the significance and impact of the inverse problem in contemporary physics. However, despite over a century of research, with seminal works by Ornstein and Zernike \cite{ornstein_integral_1914, percus_analysis_1958, percus_approximation_1962}, Yvon, Born, and Green \cite{born_kinetic_1947}, Schommer \cite{schommers_pair_1983}, and Lyubartsev and Laaksonen \cite{lyubartsev_calculation_1995}, there is surprisingly little to no evidence that these techniques can reliably extract force field parameters from experimental scattering data \cite{toth_interactions_2007}. Furthermore, there is growing evidence that existing force fields provide inaccurate representations of fluid structure when compared to experimental estimates \cite{amann-winkel_x-ray_2016, fheaden_structures_2018}. With the advent of state-of-the-art diffractometers and the rise of machine learning and high-performance computing for robust uncertainty quantification, it is relevant to revisit and contextualize prior and current work to better understand how to resolve this longstanding challenge.

Serious attempts at determining the interatomic potential from experimental scattering data began in the 1950's. Henshaw (1958) \cite{henshaw_atomic_1958} and later Clayton and Heaton (1961) \cite{clayton_neutron_1961} speculated that the ratio between the atomic collision radius and first solvation shell radius was related to the approximate width of the interatomic potential bowl. While this concept cannot directly extract the interatomic potential from the radial distribution function, it was used to conclude that argon and krypton could be reasonably represented by a (12-6) Lennard-Jones potential. Weeks, Chandler, and Anderson then introduced a separation of the pair potential into repulsive and attractive parts, in which they concluded that the repulsive part alone produces structure factors nearly identical to the repulsive and attractive parts taken together \cite{weeks_role_1971}. Henderson (1974) then proved that for a pairwise additive and homogeneous system with equal radial distribution functions that the effective interatomic potential was unique up to an additive constant \cite{henderson_uniqueness_1974}, which was later implemented numerically by Schommers (1983) \cite{schommers_pair_1983} to study liquid gallium. Around the same time, Levesque (1985) \cite{levesque_pair_1985} proposed a modified hypernetted chain closure to the Ornstein-Zernike integral relation to calculate interatomic potentials for liquid aluminum with fast convergence. Both studies were highly influential in the study of liquid metals, but offered little in resolving the inverse problem in general since interatomic potentials derived from these methods were only shown to accurately reproduce the diffusion coefficient and not other thermodynamic properties. 

The most recent inverse methods applied to experimental data are Soper's (1996) \cite{soper_empirical_1996} empirical potential structure refinement (EPSR) and Lyabartsuv and Laaksonen's (1995) \cite{lyubartsev_calculation_1995, toth_determination_2001, toth_iterative_2003} inverse Monte Carlo (IMC). EPSR is an iterative potential refinement method that is primarily used to determine real-space structures consistent with reciprocal space scattering data in fluid and glass systems. However, Soper's work on liquid water revealed that EPSR could not be reliably implemented to determine pair interaction potentials for molecular simulation applications \cite{soper_tests_2001}. On the other hand, IMC methods have been widely adopted for coarse-graining, in which the number of degrees-of-freedom of a molecular model are reduced by mapping atomic coordinates to "beads" of atom clusters. While both methods have attracted significant research interest in recent years, with the creation of an improved EPSR software package \cite{youngs_dissolve_2019} and applications of IMC in complex biological systems \cite{lyubartsev_systematic_2010} such as DNA \cite{sun_bottom-up_2021} and nucleosomes \cite{sun_bottom-up_2022}, the extraction of reliable and transferable interatomic potentials from experimental scattering data remains widely under-reported and unresolved, even for simple fluids such as noble gases. 

We recently proposed structure-optimized potential refinement (SOPR) \cite{shanks_transferable_2022} as an alternative approach to extract pair potentials from scattering data (2022). SOPR is a probabilistic iterative Boltzmann inversion (IBI) algorithm that uses Gaussian process regression to address challenges such as numerical instability and over-fitting to uncertain experimental data. SOPR derived potentials have demonstrated remarkable accuracy in predicting both the structural correlation functions and vapor-liquid equilibria of noble gases. Furthermore, the short-range repulsive decay rate determined with SOPR coincides with predictions from an independently optimized ($\lambda-6$) Mie force field for vapor-liquid equilibria \cite{mick_optimized_2015}. This finding represents the most complete example of the inverse problem in real systems and highlights the potential of scattering data in studying both macroscopic thermophysical properties of liquids and improved accuracy of local structural predictions.    

The transferability of SOPR potentials raises intriguing questions regarding what factors are most important to accurately extract local forces from scattering data. One possible explanation is that the reliability of structure inversion techniques hinges on the quality of the experimental scattering data \cite{soper_uniqueness_2007}. Levesque and Verlet (1968) speculated that experimental scattering error of $<1\%$ was required to determine the interaction potential within an error of $10\%$ \cite{levesque_note_1968}, but ultimately concluded that it is not possible to obtain quantitative information on the potential from scattering data due to systematic error in the experiments. However, it is currently unclear to what extent these previous attempts have been impeded by experimental uncertainty. The reason for this knowledge gap is that rigorous uncertainty quantification and propagation (UQ/P) is computationally demanding and requires the use of machine learning surrogate models and advanced sampling methods \cite{angelikopoulos_bayesian_2012,kulakova_experimental_2017,shanks_accelerated_2024} that were not available to liquid state theorists when these questions were first investigated. To test the hypothesis that neutron instrument accuracy is essential in force field extraction therefore lies at the crossroads of theoretical statistical physics, machine learning, and high-performance computing. 

\begin{figure}[H]
    \centering
    \includegraphics[width = 12 cm]{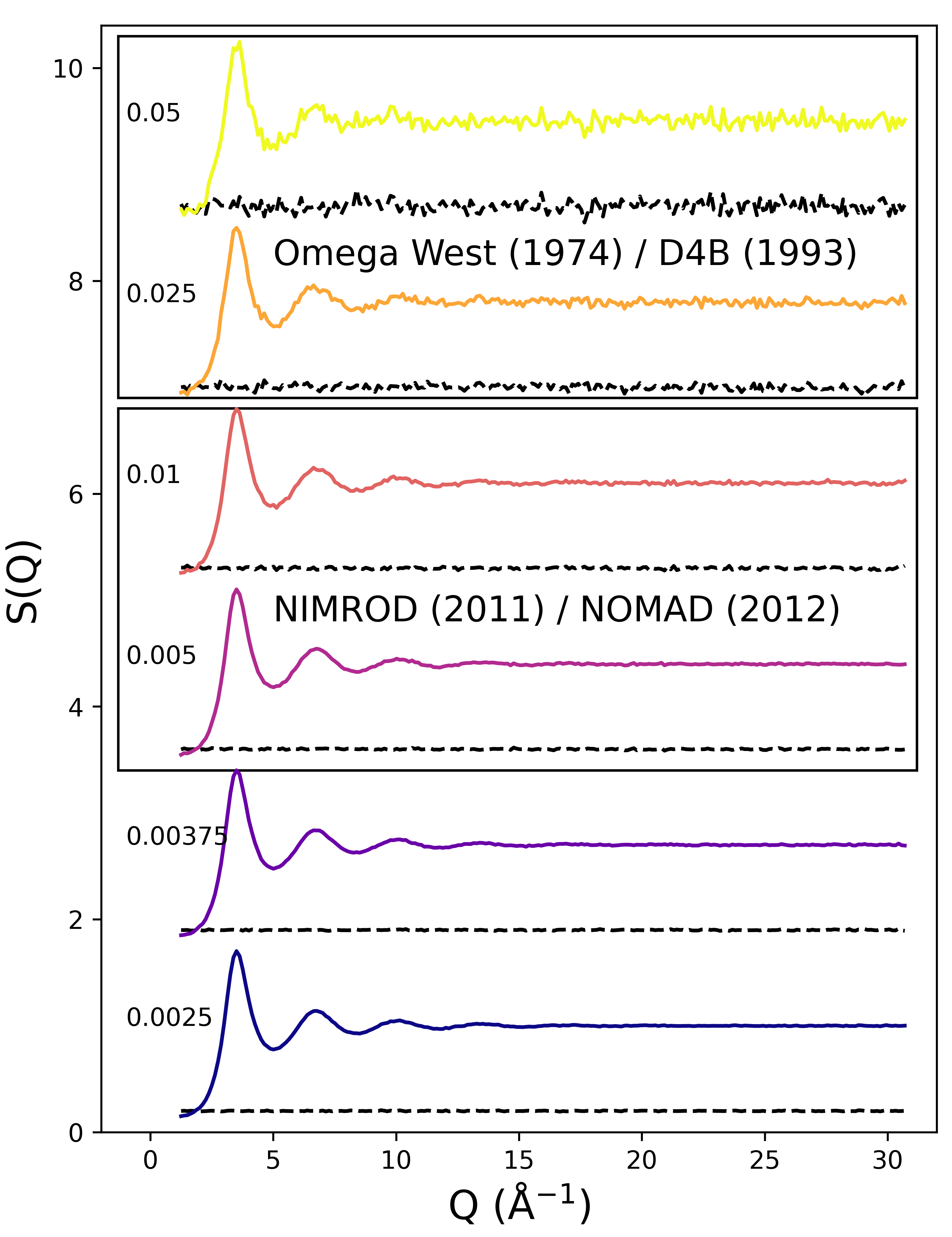}
    \caption{Static structure factors (colored lines) with introduced uncertainty (dotted black lines) for uniformly distributed noise. Measurement standard deviations $\delta S$ are labeled to the left of the structure factor.}
    \label{fig:corrupt}
\end{figure}

Here we assess how scattering measurement uncertainty impacts our ability to learn interatomic forces using a dataset of \textit{in silico} experimental structure factors with varying levels of noise. Bayesian optimization with a local Gaussian process (LGP) surrogate model was then applied to extract the underlying probability distributions on the force field parameters. Gaussian noise was introduced to a reduced Mie model structure factor with standard deviation $\delta S$, corresponding to data collected on various neutron instruments from 1973-2022. Note that the structure factor, and consequently $\delta S$, are unit-less quantities.  Constant noise at six different standard deviations, consistent with a reactor source neutron instrument \cite{willis_experimental_2017}, spanning from low to high uncertainty was added to a model structure factor (Figure \ref{fig:corrupt}). 

By studying the parameter posterior distributions as a function of introduced uncertainty, we aimed to challenge the assertion that structure factors are insensitive to the detailed form of interatomic interactions \cite{jovari_neutron_1999}. Using the ($\lambda$-6) Mie parameter Bayesian posterior distributions, we quantify how interatomic interactions such as short-range repulsion, excluded volume, and dispersion energy affect measured structure factors, shedding light on the intricate relationship between pair potentials and structural features. Surprisingly, we find that the conclusions from prior literature stating that details of the interatomic interaction could not be extracted from experimental structure factors were likely justified given the data quality available at the time, but that modern neutron instruments exceed a precision threshold where this conclusion could be overturned. These findings suggest that experimental inverse techniques were prematurely abandoned and should be revisited.

According to our results, neutron scattering measurements determined within a standard deviation of $0.005$ to a $Q_{max}\sim 30$ $\si{\angstrom}^{-1}$ are sufficient for force parameter recovery. Fortunately, this level of precision is already available at modern diffractometers, such as the Nanoscale Ordered MAterials Diffractometer (NOMAD) \cite{neuefeind_nanoscale_2012} or at other modern instruments for sufficiently long run times. Method advancements in structure inversion, along with the improvement of neutron facilities and measurement accuracy, may therefore be the key to unlock a wealth of opportunities for improving molecular models, characterizing local atomic forces, and understanding the dynamics of atoms and molecules in relation to complex and emergent physical phenomena. 

\section{Computational Methods}

In this study, we aimed to model how uncertainty propagates from neutron scattering data to the estimation of force field parameters. The impact of measurement uncertainty was isolated by constraining the Bayesian analysis to a classical model fluid. While real physical systems behave quantum mechanically and are inherently many-body in nature, classical pairwise additive model fluids continue to be studied due their low computational cost and accurate predictions of complex thermodynamic properties. Furthermore, our prior work has shown that SOPR potentials exhibit potential corrections consistent with quantum mechanical calculations \cite{shanks_transferable_2022}, suggesting that effective pair interactions could be found that capture many-body and quantum mechanical contributions.

The ($\lambda$-6) Mie fluid model was selected since it is a flexible and widely successful classical model with numerous existing and developing applications for materials modeling \cite{mick_optimized_2015, mick_optimized_2017}. The pairwise, non-bonded potential energy term of the ($\lambda$-6) Mie fluid is,

\begin{equation}
    v^{Mie}_2(r) = \frac{\lambda}{\lambda-6}\bigg(\frac{\lambda}{6}\bigg)^{\frac{6}{\lambda-6}} \epsilon \bigg[ \bigg(\frac{\sigma}{r}\bigg)^\lambda - \bigg(\frac{\sigma}{r}\bigg)^6 \bigg]
\end{equation}

\noindent where $\lambda$ is the short-range repulsion exponent, $\sigma$ is the collision diameter (distance), and $\epsilon$ is the dispersion energy (energy) \cite{mie_zur_1903}.

\subsection{Modeling Neutron Measurement Uncertainty in a Mie Fluid Model}

To model experimental uncertainty, a set of Mie fluids was simulated with sufficient sampling statistics to calculate a highly-accurate static structure factor ($\delta S(Q) < 0.001$) to $Q_{max} = \sim 30$ \AA$^{-1}$). Computer generated atomic trajectories were calculated in HOOMD-Blue \cite{anderson_hoomd-blue_2020}. MD simulations were initiated with a random configuration of 500 particles at reduced density $\rho^* = 0.1$ and reduced temperature $T^* = 1$ and equilibrated with Langevin dynamics for $1\times10^5$ timesteps ($dt=10$ femtoseconds). Potentials were truncated at $3\sigma$ with an analytical tail correction, and radial distribution functions were calculated with Freud \cite{ramasubramani_freud_2020}. Static structure factors are calculated via radial Fourier transform of the radial distribution function. 

Experimental measurements of structure factors are subject to uncertainty arising from various factors, including experimental, model, and numerical sources. Uncertainties in neutron flux, energy, time-of-flight, minimum and maximum momentum transfer ($Q_{min}$, $Q_{max}$), and data collection time contribute to uncertainty in neutron counting statistics and the effective resolution of the instrument. Post-processing corrections for inelastic, incoherent, multiple, and low momentum transfer scattering further contribute to uncertainty in the structure factor form \cite{soper_inelasticity_2009}. These effects result in variations in the neutron intensity that are not necessarily normally distributed \cite{barron_entropy_1986}; however, errors from neutron detection are normal due to the limiting behavior of the Poisson distribution for large number of counts \cite{heybrock_systematic_2023}. For reactor source instruments, the variance due to these random errors remains approximately constant to a limited $Q$-max (10-20 $\si{\angstrom}^{-1}$), while for spallation sources, the variance increases proportionally to the square of the momentum transfer to a higher $Q$-max of 50-125 $\si{\angstrom}^{-1}$ \cite{neuefeind_nanoscale_2012}. Currently, the extent to which this uncertainty influences the accuracy and reliability of force field reconstructions remains unknown.  

Uncertainty quantification was performed for reactor type neutron instruments by adding Gaussian noise with standard deviations equal to twice the values of those indicated in Figure \ref{fig:corrupt} to four replicates of the simulated structure factor. Using multiple replicates of the noisy structure factor as a data target reduces the chance of over fitting to a single, randomly generated structure factor. Of course, this modeling approach is not directly consistent with an actual scattering measurement which is typically reported as a single measurement over a fixed length of time. However, it is well-known that the standard deviation in neutron counting statistics is proportional to the square-root of the number of counts \cite{heybrock_systematic_2023}. Assuming that neutron counts are equally distributed over the course of a measurement, we then expect that four structure factor replicates with twice the standard deviation of the target structure factor is approximately equivalent to the single target scattering pattern. 

Bayesian analysis was then performed over 16 \textit{in silico} experimental conditions on a $4 \times 4$ equal spaced grid with $\sigma = [1.85, 1.89, 1.93, 1.97]$, $\epsilon = [0.86, 0.80, 0.74, 0.70]$ and fixed $\lambda=12$. Since spallation type neutron instruments have a $Q^2$-dependent random error, UQ on the constant error can be interpreted as an uncertainty upper bound for spallation type instruments. Uncertainty levels were selected based on published data of structure factors measured on neutron instruments from the early 1970's to 2022. Notably, a classic argon data set collected at the Omega West reactor (1973) \cite{yarnell_structure_1973} is well approximated by constant noise with variations in $S(Q)$ of approximately $\delta S = 0.05$, as noted in Figure \ref{fig:corrupt}. Similarly, krypton data collected on D4B (1993)\cite{barocchi_neutron_1993} is approximated by the constant $\delta S = 0.025$ case, while modern instruments such as NOMAD and NIMROD (>2010) exhibit uncertainty distributions similar to the $\delta S = 0.005-0.01$ cases depending on the data collection time \cite{bowron_nimrod_2010, neuefeind_nanoscale_2012}. Two low uncertainty extremes were chosen beyond these reported values to identify measurement precision thresholds and model trends in the predictability of force parameters.  

\subsection{Bayesian Uncertainty Quantification for Force Field Reconstruction}

According to the Henderson inverse theorem, it is theoretically possible to uniquely recover the underlying potential in a pairwise additive, homogeneous fluid \cite{henderson_uniqueness_1974}. In the context of Bayesian optimization, Henderson's theorem requires that there should be a global maximum in the posterior probability distribution at the unique force parameters. However, as the uncertainty in the structure factor signal increases, deviations from this unique potential are expected, causing the probability distributions to broaden. This broadening indicates a decrease in confidence in the estimation of the model parameters. In other words, as structural uncertainty increases, our ability to accurately predict the potential energy decreases, leading to a wider range of possible parameter values to explain the data. 

Bayesian inference was implemented to calculate parameter probability distributions as a function of structure factor uncertainty. For simplicity of notation, let $\boldsymbol{\theta} = \{\lambda, \sigma, \epsilon, \sigma_n\}$ represent the model parameters and $\mathcal{Y} = S_d(Q)$ be the structure factor observations. The nuisance parameter, $\sigma_n$, represents the width of the Gaussian likelihood and captures uncertainty from the experimental data and Gaussian process model, which is not known \textit{a priori}. Calculating the posterior probability distribution with Bayesian inference then requires two components: (1) prescription of prior distributions on the model parameters, $p(\theta)$, and (2) evaluation of the structure factor likelihood, $p(\mathcal{Y}|\boldsymbol{\theta})$. The prior distribution over the $(\lambda-6)$ Mie parameters is assumed to be a multivariate log-normal distribution,

\begin{equation}
     \theta - \gamma_\theta \sim \log \mathcal{N} (\mu_{\boldsymbol{\theta}} + \gamma_\theta, \sigma^2_{\boldsymbol{\theta}})
\end{equation}

\noindent where $\mu_{\boldsymbol{\theta}}$ and $\sigma^2_{\theta}$ are the prior mean and variance of each parameter in $\boldsymbol{\theta}$ and $\gamma_\theta \in \mathbb{R}$ is a real-valued parameter shift that enforces a lower bound. A wide, shifted multivariate log-normal distribution was selected because it is non-informative and imposes non-negativity constraints on the model parameters. Specifically, $\lambda-6$ (defined by the Mie type fluid), $\sigma$, $\epsilon$, and $\sigma_n$ must be positive. For reference, the prior parameters used in this study and sample range is summarized in Table \ref{tab:priors}.

\begin{table}[H]
\centering
\caption{\label{tab:priors}
Prior parameters on the ($\lambda$-6) Mie model parameters.}
\begin{tabular}{| c | r | r | r |}
\hline
\textrm{Parameter}&
\textrm{$\mu$}&
\textrm{$\sigma (std.)$}&
\textrm{$\gamma_\theta$}\\
\hline
$\lambda$  &  3    & 1 & 6 \\
$\sigma$   &  2    & 1 & 0\\
$\epsilon$ &  0.7  & 1 & 0\\
$\sigma_n$ &  0.1  & 3 & 0\\
\hline
\end{tabular}
\end{table}

The likelihood function is a Poisson distribution of the neutron counts, but we can approximate this distribution as a Gaussian because a Poisson distribution approaches a Gaussian distribution in the high count limit,

\begin{equation}\label{eq:gausslikelihood}
    p(\mathcal{Y}|\boldsymbol{\theta}) = \bigg(\frac{1}{\sqrt{2 \pi}\sigma_n}\bigg)^\eta \exp\bigg[-\frac{1}{2\sigma^2_{n}}\sum_j\ [{S}_{\boldsymbol{\theta}_i}(Q_j) - S_d(Q_j)]^2\bigg]
\end{equation}

\noindent where ${S}_{\boldsymbol{\theta}}(Q_i)$ is the molecular simulation predicted structure factor, $\eta$ is the number of observed points in the structure factor, and $j$ indexes over these points along the momentum vector. Bayes' theorem is then expressed as,

\begin{equation}\label{eq:inference}
    p(\boldsymbol{\theta}|\mathcal{Y}) \propto p(\mathcal{Y}|\boldsymbol{\theta}) p(\mathbf{\boldsymbol{\theta}})
\end{equation}

\noindent where equivalence holds up to proportionality. This construction is acceptable since the resulting posterior distribution can be normalized \textit{post hoc} to find a valid probability distribution. For further details see the following excellent reviews of Bayesian inference \cite{gelman_bayesian_1995, bishop_pattern_2006}. 

The Bayesian likelihood distribution is estimated using Markov Chain Monte Carlo (MCMC) samples over the model parameters $\theta = \{\lambda, \sigma, \epsilon, \sigma_n\}$. Computationally, a sample of the model parameters is drawn from a Metropolis-Hastings type algorithm, passed to the surrogate model, evaluated, and compared to the \textit{in silico} structure factor. MCMC samples were calculated using the emcee package \cite{foreman-mackey_emcee_2013} from 160 walkers with dynamic burn-in and sample time based on the autocorrelation convergence criterion used in the emcee package and default stretch move with dynamic tuning. 

After the posterior distributions were computed for all experimental conditions and uncertainty levels, the marginal posterior histograms were averaged over all experimental conditions. Histogram averaging was performed by defining a fixed range and bin count for all posterior marginals and weighting the distribution counts by the ratio of the number of counts for a given experiment with the total number of counts over all experiments. In other words, if $p_{a,b}(\theta)$ is the marginal posterior probability distribution for experiment condition $a$, uncertainty $b$, on model parameter $\theta$, then the average marginal posterior $\mathcal{P}_b(\theta)$ for uncertainty level $b$ is given by,

\begin{equation}
    \mathcal{P}_b(\theta) \approx \sum_{a} \frac{n_{a,b}}{\sum_{a}n_{a,b}}p_{a,b}(\theta)
\end{equation}

\noindent where $n_{a,b}$ is the number of independent MCMC samples for experiment $a$ and uncertainty $b$. Conceptually, these average marginal posteriors are an approximation to the marginal parameter posterior distributions over the joint probability density containing the model parameters, structure factor, and thermodynamic state, $p(\lambda,\sigma,\epsilon,S(Q), T^*, \rho^*)$, where $T^* = k_bT/\epsilon$ and $\rho^* = \rho \sigma^3$ are the reduced temperature and density, respectively. More rigorous approximations derived from methods such as Gibbs sampling were not implemented due to the extremely high computational cost per experiment.

\subsection{Local Gaussian Process Surrogate Models for Structure Factors}

The process of populating the posterior distribution function necessitates the evaluation of likelihood for each condition of interest within the model parameter space, which, in turn, requires conducting an infeasible number of molecular dynamics simulations. The computational burden associated with this procedure renders the Bayesian framework impractical even for a relatively small number of samples. To illustrate this, consider the task of obtaining a collision diameter posterior distribution with a grid resolution of approximately 2\% across a wide prior range of $m_{\sigma}$ (ranging from 0.5 to 1.5). Achieving such resolution for just the $m_{\sigma}$ parameter alone would demand a minimum of 50 samples. If the same level of resolution is desired for the remaining two parameters, a staggering 125,000 molecular simulations are required to comprehensively quantify the posterior distribution space. Clearly, there is a substantial computational challenge involved in obtaining accurate and comprehensive posterior distributions within the Bayesian framework.   

To expedite the evaluation of the Bayesian likelihood, a local Gaussian process (LGP) surrogate model was trained to generate structure factors based on a training set of 960 randomly sampled ($\lambda$-6) Mie parameters in a prior range specified in Table \ref{tab:ranges} \cite{shanks_accelerated_2024}. This range of parameters was selected to correspond with the liquid phase region of the Mie phase diagram and avoid pathological simulations that can occur near phase transitions. Note that this range will change based on the arbitrary choice of temperature and density for the Mie fluid simulation; but, since the reduced phase diagram is simply scaled from these values, the molecular dynamics simulation will have the same dynamics, average thermodynamic properties and structure.

\begin{table}[H]
\centering
\caption{Estimated boundaries for physics-constrained prior space based on the ($\lambda$ - 6) Mie fluid phase diagram. $m = 6$ is the attractive tail exponent of the ($\lambda$ - 6) Mie potential. $*$) The maximum $\lambda$ was selected to be substantially larger than previously reported values \cite{vrabec_set_2001, mick_optimized_2015, shanks_transferable_2022}.}
\begin{tabular}{| c | l | l | l | l |}
\hline
\textrm{Param.}&
\textrm{Min.}&
\textrm{Min. Criteria}&
\textrm{Max.}&
\textrm{Max. Criteria}
\\
\hline
$\lambda$  & 6.1  & $m=6 \implies \lambda>6$  &  18 & Literature$^*$ \\
$\sigma$   & 1.5 & Vapor-Liquid Equil. &  2 & Solid-Liquid Equil.  \\
$\epsilon$ & 0.2 & $\epsilon<0$ undefined  &  1.1 & Vapor-Solid Equil. \\
\hline
\end{tabular}
\label{tab:ranges}
\end{table}

LGP surrogate models reduce the computational time complexity of standard GP regression with little loss in predictive accuracy \cite{das_block-gp_2010, chen_parallel_2013, gramacy_local_2015}. Hyperparameter selection was performed using Bayesian optimization with Markov chain Monte Carlo and surrogate model accuracy was determined to have a root-mean-square error (RMSE) of 0.0036 for a 160 randomly sampled test set within the surrogate parameter range. Surrogate model validation and hyperparameter training is discussed further in the Appendix.

\section{Results and Discussion}

We first explore parameter sensitivity using the analytical derivative of the LGP surrogate model. The surrogate model derivative can quantify how changes in each of the Mie parameters affect the structure factor, providing detailed insights into structural behavior under various model conditions. Thanks to the computational efficiency of the LGP surrogate, it is now possible to construct sensitivity heat maps across the ($\lambda$-6) Mie fluid phase diagram.

Our second result involves performing Bayesian analysis on structure factors derived from a grid of 16 reference Mie parameters, each subjected to 6 different noise levels. The objective was to determine the critical threshold of scattering precision necessary to extract force field parameters from structure factors. Once the threshold accuracy was identified, posterior distributions at this critical noise level were analyzed to estimate the expected credibility intervals on force parameter recovery in simple liquids.

\subsection{Sensitivity Analysis with Local Gaussian Process Derivatives}

The LGP surrogate model can quantify the impact of varying ($\lambda$-6) Mie parameters on the structure at specific $Q$ values via estimation of its derivative with respect to the model parameter, $\frac{\partial S(Q)}{\partial \theta}$, excluding the nuisance parameter $\sigma_n$. Zeros and extrema of the LGP derivative reveal regions of the structure factor that are least and most affected by changes in the force parameters, respectively.

\begin{figure}[H]
    \centering
    \includegraphics[width = 16cm]{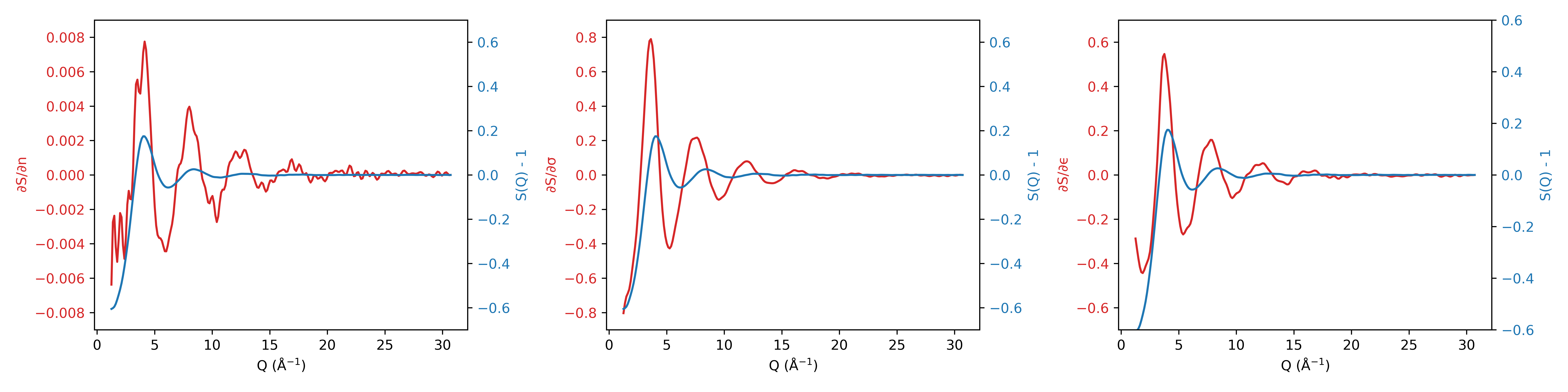}
    \caption{Derivatives of the surrogate model-predicted structure factor, S(Q), with respect to each parameter of the ($\lambda-6$) Mie force field (red line) plotted with the given structure factor (blue line).}
    \label{fig:derivative}
\end{figure}

Figure \ref{fig:derivative} illustrates the derivatives of the surrogate model-predicted structure factor with respect to each parameter of the ($\lambda-6$) Mie force field. Even small changes in the Mie parameters result in substantial modification of the structure factor patterns. Consequently, provided experimental scattering results meet a necessary accuracy threshold, all three of the Mie model parameters (short-range repulsion, size, and dispersive attraction) could be identified. Changes to the structure factor; however, do not impact all force parameters equally. The repulsive exponent derivative exhibits a small magnitude and undergoes sign changes near the full-width half maximum of the structure factor peaks. This behavior suggests that increasing the repulsive exponent, which determines the "hardness" of the particles, causes a slight increase in height and narrowing of the structure factor peaks without significantly affecting their location. In the case of the collision diameter, zeros of the derivative occur at structure factor peaks and troughs, while local extrema align with the half-maximum positions. Consequently, increasing the effective particle size shifts the structure factor towards lower $Q$ values while maintaining relatively constant peak heights. Regarding the dispersion energy, its derivative displays zeros at the half-maximum positions of the structure factor and local extrema near the peaks and troughs. This behavior indicates that an increase in the dispersion energy leads to an increased magnitude and sharpening of the structure factor peaks similar to the repulsive exponent.

\begin{figure}[H]
    \centering
    \includegraphics[width = 16cm]{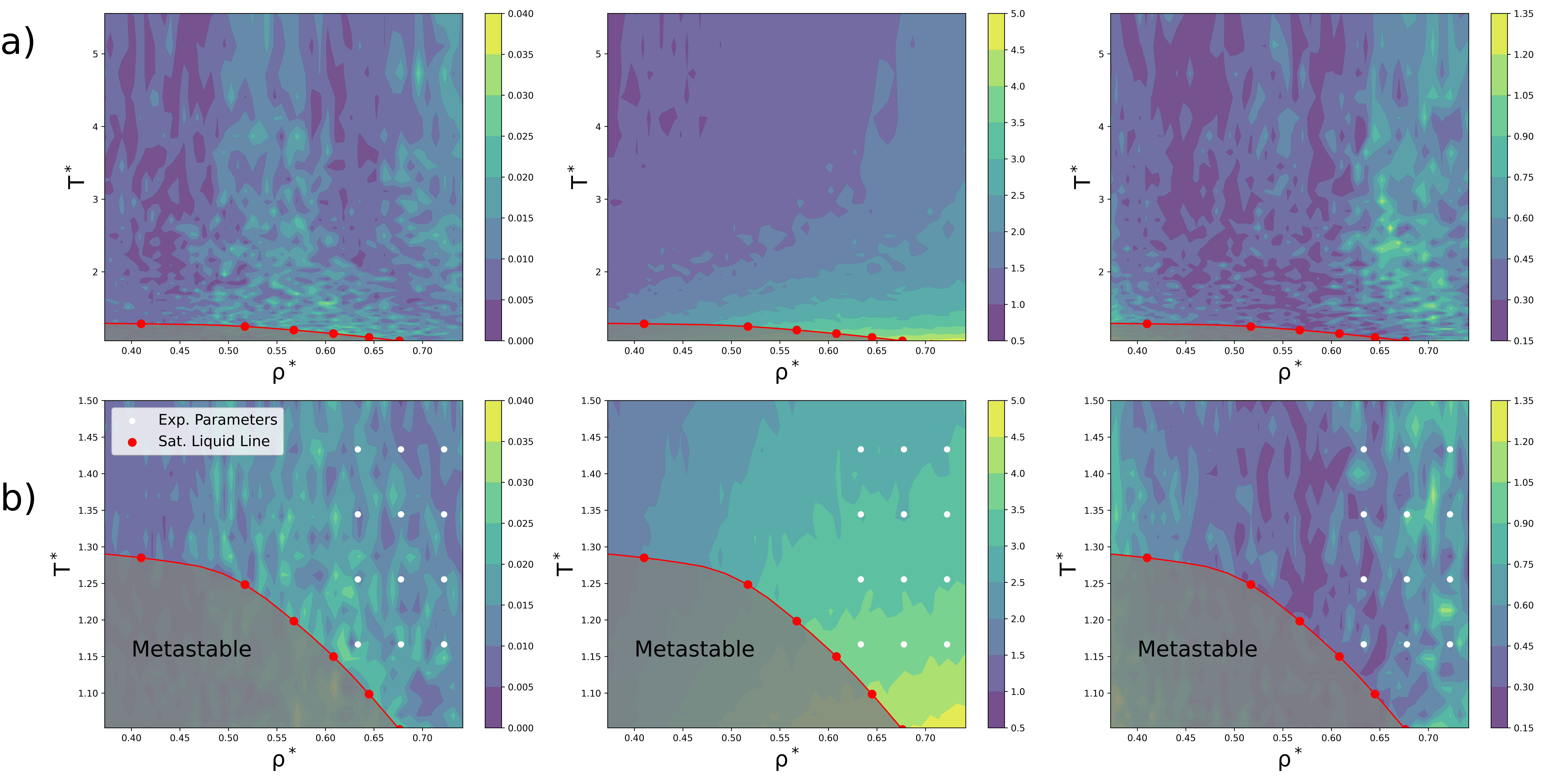}
    \caption{(a) Heat map of the maximum absolute value of the structure factor derivative with respect to each parameter of the ($\lambda-6$) Mie force field for the validated range of the surrogate model and (b) near the vapor-liquid coexistence line (red line). The grey region represents a metastable fluid that separates into vapor and liquid phases.}
    \label{fig:derphasediag}
\end{figure}

Computation of the LGP derivative can also be performed over the entire validated range of the surrogate model (see Appendix). To visualize the results, we present a heat map of the maximum of the absolute value of the derivative with respect to reduced temperature $T^* = T/\epsilon$ and reduced density $\rho^* = \rho \sigma^3$ in Figure \ref{fig:derphasediag}. Higher values (yellow regions) indicate a high sensitivity of the structure factor relative to lower values (blue regions). First note that the maximum derivative estimates vary in magnitude significantly, with a two orders-of-magnitude smaller value for the repulsive exponent (0.04) compared to the collision diameter (5.0) and dispersion energy (1.35). The repulsive exponent $\lambda$ exhibits biomodality as a function of $\rho^*$, with higher sensitivity regions tending towards higher densities. The structural sensitivity at high density could be explained by the fact that such systems tend to collide more frequently at close range which is where the repulsive exponent strongly influences the potential energy function. The collision diameter $\sigma$ has a clear trend with more sensitive regions being higher density and lower temperature. Observing the sensitivity over the full range (Figure \ref{fig:derphasediag}a), we can see that there also appears to be asymptotic behavior near specific densities, suggesting that higher sensitivities correspond to closer proximity of the atoms where excluded volume effects can dominate the structure. Finally, the dispersive forces appear to become more significant to the structure near and above the critical temperature and at high density and temperatures. Of course, to fully elucidate trends and patterns in structural sensitivity to the interaction potential would require an LGP surrogate model trained over a wider space on the Mie fluid phase diagram.

\subsection{Force Field Parameter Posterior Distributions as a Function of Uncertainty}

\begin{figure}[H]
    \centering
    \includegraphics[width = 16 cm]{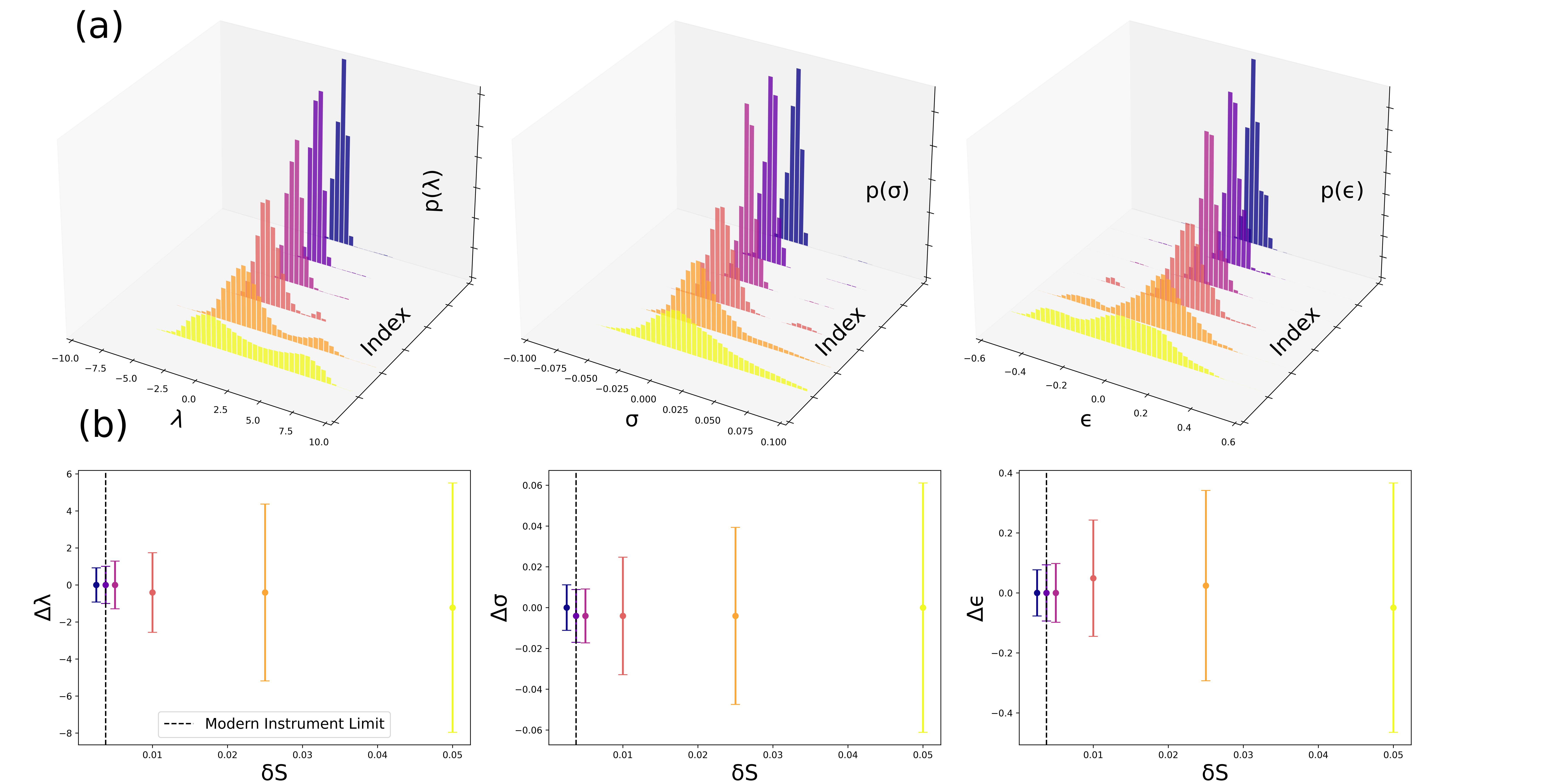}
    \caption{Average marginal distributions computed from the 16 reference posterior distributions. (a) Histograms of the average marginal distributions on the ($\lambda-6$) Mie force field parameters as a function of uncertainty in the structure factor ($\delta S$). (b) MAP estimates (points) are plotted with 2 std. dev. error bars as a function of noise. Low parameter uncertainty cases (blue) are compared to high uncertainty cases (red) and the lower limit precision of current neutron instruments (black dashed line). Low and high uncertainty distributions were separated based on the near doubling of the standard deviation between $\delta S = 0.005$ and $\delta S = 0.01$.}
    \label{fig:marginals}
\end{figure}

Armed with a precise and fast surrogate model for the Mie fluid structure factor, we can now proceed to evaluate the likelihood function and, consequently, derive Bayesian posterior distributions. Figure \ref{fig:marginals} illustrates experiment averaged marginal probability distributions of $(\lambda, \sigma, \epsilon)$ as a function of noise. 

The 1D Marginal distributions in Figure \ref{fig:marginals} are obtained by integrating the joint posterior probability distribution over all but one model parameter. The mode of the marginal distribution corresponds to the marginal \textit{maximum a posteriori} (\textit{MAP}). It is worth noting that as the uncertainty in the structure factor increases, the marginal distributions become wider and flatter. This behavior is expected, as greater uncertainty in the observation leads to increased uncertainty in the parameter distribution. In cases where the structure factors exhibit low uncertainty, the \textit{MAP} estimates accurately recover the unknown force field parameters. Deviation between the \textit{MAP} estimate and true parameter value is calculated as a function of uncertainty and presented in Table \ref{tab:maps}. 

\begin{table}[H]
\centering
\caption{\label{tab:maps} Error in ($\lambda-6$) Mie force field parameters determined by Bayesian inference on the structure factor. $\Delta p$ is the difference between the \textit{MAP} estimate and the underlying parameter set.}
\begin{tabular}{| c | r r r r |}
\hline
\textrm{Comparable Neutron Instrument}&
\textrm{$\delta S$}&
\textrm{$\Delta \lambda \pm 2\sigma_\lambda$}&
\textrm{$\Delta \sigma \pm 2\sigma_\sigma$}&
\textrm{$\Delta \epsilon \pm 2\sigma_\epsilon$}\\
\hline
-                                               & 0.0025  & -0.051 $\pm$ 0.9 &  0.000  $\pm$ 0.01 &  0.00 $\pm$ 0.07 \\
-                                               & 0.00375 & -0.255 $\pm$ 1.0 & -0.002  $\pm$ 0.01 &  0.00 $\pm$ 0.09 \\
NOMAD (2012) \cite{neuefeind_nanoscale_2012}    & 0.005   &  0.051 $\pm$ 1.3 & -0.002  $\pm$ 0.01 &  0.02 $\pm$ 0.10 \\
\hline
NIMROD (2010) \cite{bowron_nimrod_2010}         & 0.01    & -0.561 $\pm$ 2.1 & -0.006  $\pm$ 0.02 &  0.03 $\pm$ 0.19 \\
D4B (1993) \cite{barocchi_neutron_1993}         & 0.025   & -0.561 $\pm$ 4.8 & -0.004  $\pm$ 0.03 &  0.03 $\pm$ 0.29 \\
Omega West (1973) \cite{yarnell_structure_1973} & 0.05    & -1.173 $\pm$ 6.7 & -0.002  $\pm$ 0.04 & -0.05 $\pm$ 0.36 \\
\hline
\end{tabular}
\end{table}

First, note the drastic difference in the accuracy of the \textit{MAP} estimates for the repulsive exponent and dispersion energy parameters as we transition from an uncertainty level of $\delta S = 0.025$ to $\delta S = 0.05$. The $\sigma$ parameter is accurately estimated in both scenarios, demonstrating its reliable prediction even for low quality scattering data. In the $\delta S = 0.025$ case, the $\lambda$ and $\epsilon$ parameters are also accurately predicted. However, for the $\delta S = 0.05$ case, the $\lambda$ and $\epsilon$ parameters become unlearnable with \textit{MAP} deviations of -1.173 and -0.05, respectively. 

The data also shows a significant change in the width of the distributions at critical uncertainty levels. The standard deviation effectively doubles for the Mie parameters between $\delta S = 0.005$ and $\delta S = 0.01$. This rapid increase in width of the posterior distribution is significant since it becomes exceedingly more difficult to estimate the potential parameters using optimization techniques. Based on these shifts in the standard deviations, we recommend that neutron scattering experiments for liquids not exceed random errors of $\delta S = 0.005$ to $\sim$ 30 \AA$^{-1}$ if attempting to extract pair potential information from the structure factor. This level of precision is achievable on modern instruments but may require longer run times than standard neutron scattering measurements.

Taken together, these observations are critical to contextualizing prior studies in which it has been concluded that the structure factor is insensitive to the interatomic interactions beyond the excluded volume \cite{jovari_neutron_1999, hansen_theory_2013} or that uncertainty in the structure measurement impeded prediction of transferable potentials \cite{schommers_pair_1983,levesque_pair_1985,lyubartsev_calculation_1995}. In these studies, the instrument uncertainty ranged from values of 0.03-0.07, exceeding the precision threshold identified by our model. Barocchi's (1993) study on liquid krypton was unique in the conclusion that the neutron instrument accuracy was now high enough to elucidate detailed many-body interactions \cite{barocchi_neutron_1993}, which was based on structure factors measured to a precision $\leq 0.025$, which is consistent with our conclusions.

\subsection{Uncertainty Quantification on a State-of-the-Art Neutron Instrument Model}

\begin{figure}
    \centering
    \includegraphics[width = 16 cm]{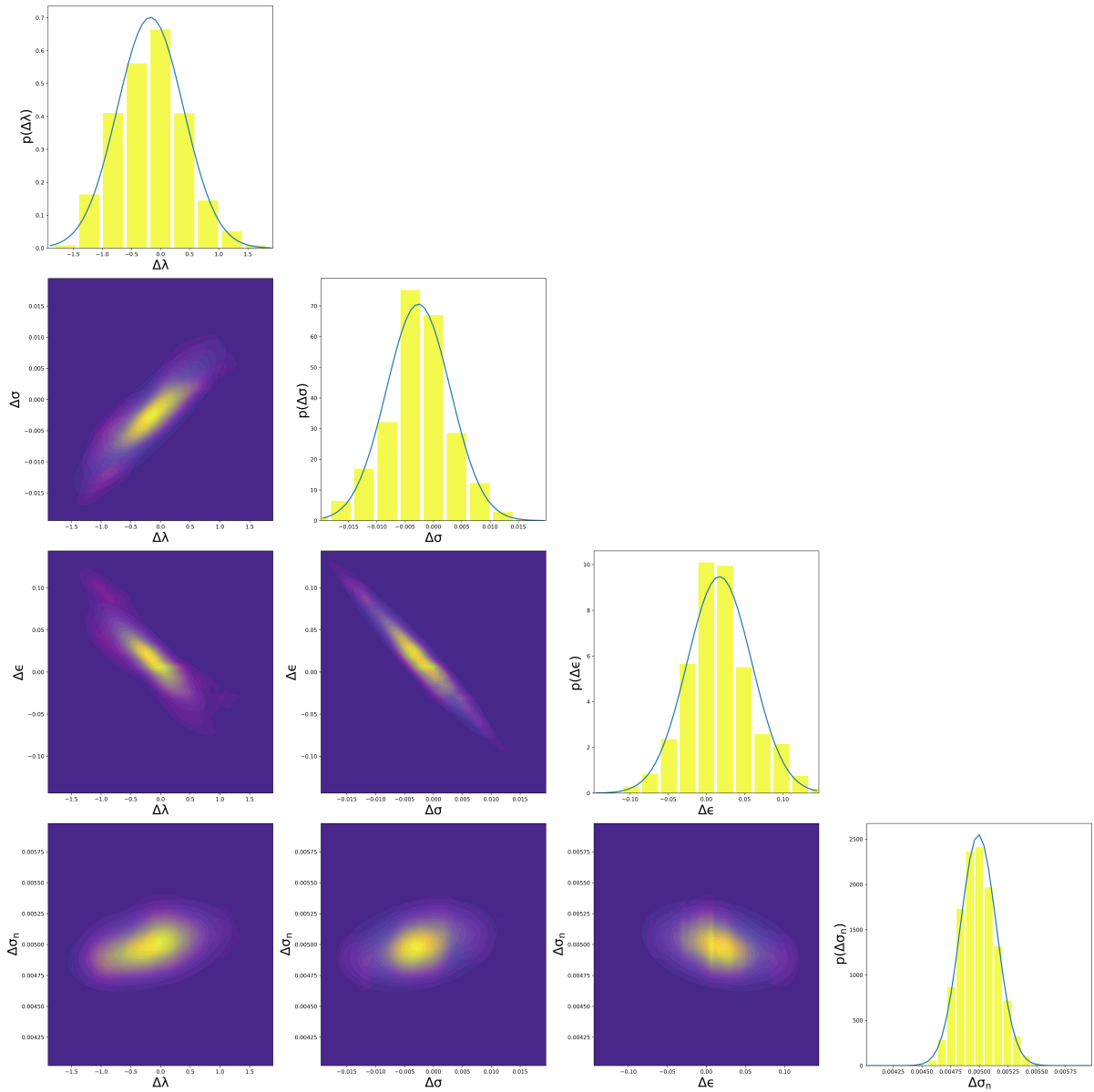}
    \caption{Marginal distributions on the ($\lambda-6$) Mie force field parameters for $\delta S = 0.005$ at 30 \AA $^{-1}$ with variance sampled with MCMC plotted with known parameter values (red dashed line).}
    \label{fig:histograms}
\end{figure}

We have demonstrated through uncertainty quantification that prior attempts of the statistical mechanical inverse problem were likely limited based on the accuracy of experimental scattering measurements. Specifically, noise in structure factor data can significantly impact the broadness of parameter probability distributions, rendering optimization methods unable to accurately estimate the parameter \textit{MAP}. Furthermore, modern diffractometers are sufficiently precise to provide reliable inverse problem solutions to assess a variety of atomic force properties. Consequently, we further analyze the posterior distributions for structure factor results that are consistent with modern diffractometers. The highest flux instruments are spallation sources, which can measure structure factors with constant standard deviations of $\delta S = 0.005$ out to 30 \AA$^{-1}$. This condition well-represents an upper bound of  uncertainty in a structure factor measurement on the state-of-the-art NOMAD and NIMROD instruments. Posterior marginals, MCMC samples, and heat maps of the joint posterior distribution are illustrated in Figure \ref{fig:histograms}.

The marginal \textit{MAP}, corresponding to the global maximum of the marginal distribution, accurately predicts the true parameter values (indicated by red dashed lines) with exceptional precision, exhibiting error rates below 1\% for all force parameters. The shape and width of the marginal distributions offer valuable insights into the influence of each parameter on the ensemble fluid structure. The collision diameter marginal exhibits a narrow and symmetric shape, characterized by a probability density at the \textit{MAP} that surpasses the repulsive exponent and dispersion energy by factors of 80 and 3, respectively. This symmetry and high probability density suggest a remarkable sensitivity of the structure factor to changes in the effective particle size which is consistent with the observations of Weeks, Chandler and Anderson \cite{weeks_role_1971}. However, the seemingly small difference between the repulsive structure factor alone and the true structure factor clearly contains sufficient information to determine the potential well-depth as well as the repulsive exponent and collision diameter. Therefore, we contend that the structure factor of liquids contains more information than previously believed. 

Two standard deviations of the posterior distribution can be used as an estimate of our confidence in recovering the force parameter from the structure factor with $\sim$95\% confidence. Using this metric, we find that the force parameters can be recovered with 95\% confidence within $\pm$1.3 for the repulsive exponent, $\pm$0.02 $\sigma$, and $\pm$0.1 $\epsilon$. Of course, the posterior distributions computed here are in reduced Lennard-Jones (LJ) units and must be scaled by a known reference to approximate the credibility intervals in real units. For example, taking the LJ parameters for argon ($\sigma = 3.4$ \AA, $\epsilon = 0.24$ kcal/mol \cite{shanks_transferable_2022}) would give a real unit estimate of $\lambda \pm$1.3, $\sigma\pm$0.068 \AA, and $\epsilon\pm$0.024 kcal/mol with 95\% confidence.

Uncertainty quantification and propagation of the potential in relation to the structure factor holds the key to unlocking several capabilities of neutron scattering, including force field design, elucidation of many-body interactions, and improved understanding of structural properties in fluid systems \cite{terban_structural_2022}. While these aims have motivated research on the inverse problem for over a century, we are only now seeing evidence that accurate structure inversion on experimental data is a possibility. We argue, despite having presented a study on a simple model, that our results warrant the recommendation of revisiting inverse methods for real fluids. 

One exciting prospect for inverse problem methods is that interaction potentials derived from structure can serve as an external validation to computationally expensive bottom-up atomistic models. One example is electron structure calculations, in which a highly accurate quantum mechanical treatment of the electron structure can reveal insights into potential energy surfaces and reaction mechanisms. Electron structure methods have become faster and more robust due to quantum computing \cite{motta_emerging_2022}, clever basis set selection \cite{ratcliff_flexibilities_2020}, and machine learning \cite{duan_learning_2019,rath_framework_2023}. As these more fundamental theories of atomic structure and motion become commonplace, experimental neutron scattering data will be a crucial validate of their predictions. Indeed, we have already shown that many-body interactions in noble gases are consistent with electron structure calculations in trimeric systems \cite{guillot_triplet_1989,shanks_transferable_2022}. However, further advancement of inverse methods can provide quantifiable validation of many-body interactions in progressively complex systems.

In contrast to experimental analysis, inverse methods for coarse-graining have thrived in contemporary chemical physics \cite{shell_relative_2008,moore_derivation_2014,rudzinski_generalized-yvon-born-green_2015,frommer_variational_2022,ivanov_coarse-grained_2023}. In coarse-graining, structure factors (or more commonly the partial radial distribution functions) are generated from a known model where structural uncertainty fluctuations are significantly smaller than that of experimental data. In pairwise additive systems with low structural uncertainty, our Bayesian analysis indicates the likely presence of global maxima in narrow posterior distributions, suggesting that optimization schemes should be capable of reliably identifying these maxima. Since global maxima are also expected in experimental scattering measurements conducted using state-of-the-art neutron instruments, there is an opportunity to employ these maximum-likelihood methods for developing novel force fields directly from experimental data. With this evidence, we hope to bridge the gap between experiment and simulation-based inverse techniques and foster closer collaboration between these two communities.

It is important to acknowledge certain limitations in the previous analysis when considering the extension of the results to other physical systems. First, the sensitivity of more complex systems than the ($\lambda$-6) Mie model may differ from the estimates reported in our study. Therefore, it is cautioned to interpret the results of this example as a conceptual exploration of how classical two-body interactions impact the atomic organization in fluids. Hence, the specific response of complex systems to variations in interatomic forces should be studied individually. Second, if a fluid cannot be adequately described by a ($\lambda$-6) Mie model, the resulting posterior distribution may exhibit flatness or multimodality, indicating a high level of uncertainty in both the structure and model parameters. In such cases, a more accurate model of the system should be adopted to facilitate reliable parameter inference. Furthermore, systematic errors were not investigated and are certainly significant to the potential reconstruction. Therefore, further work should explore how to eliminate systematic errors in neutron scattering analysis through physics-based Gaussian process regression or analogous approaches.

\section{Conclusions}

Rigorous uncertainty quantification and propagation analysis has shown that modern neutron diffractometers have attained the necessary accuracy for reliable force field reconstruction. It has also been shown that neutron scattering measurements within $\leq 0.005$ at $\sim$30 $\si{\angstrom}^{-1}$ are sufficient for force parameter recovery in simple liquids. We stress that the structure factor contains information on the force field parameters that control the attractive as well as the repulsive part of the interatomic potential. This study highlights the exciting possibility of using neutron scattering to predict the potential energy function of Mie-type fluids, emphasizes the critical role of experimental precision in extracting potentials from scattering data, and offer svaluable insights into the nature of interatomic forces in liquids. 

The significance of these results extends beyond the field of neutron scattering analysis. They hold great potential in advancing force field design and optimization, enabling the development of effective coarse-graining techniques, and facilitating the exploration of many-body effects in fluid ensembles. The far-reaching impact of machine learning-accelerated methods in predicting interatomic forces from experimental structure measurements is evident. In summary, this research establishes the transformative potential of machine learning in extracting interatomic forces from experimental structure measurements with uncertainty quantification.

\section*{Data Availability}

The datasets generated during and/or analysed during the current study are available from the corresponding author on reasonable request. Python codes for data generation and analysis are provided on GitHub at https://github.com/hoepfnergroup/Bayesian-Neutron-Potentials.

\section*{Acknowledgements}

This study is supported by the National Science Foundation Award No. CBET-1847340. The support and resources from the Center for High Performance Computing at the University of Utah are gratefully acknowledged.

\section*{Author Contributions}

B.L.S contributed towards conceptualization, algorithm development and implementation, data visualization, and writing. H.W.S provided contributions to conceptualization, algorithm development and implementation. M.P.H contributed to conceptualization, funding acquisition, and manuscript revision and preparation.

\section{Competing Interest}

The authors declare no conflicts or competing interests.

\printbibliography

\section{Appendix}
\appendix
\input{si.tex}

\typeout{get arXiv to do 4 passes: Label(s) may have changed. Rerun}

\end{document}

%% file: si.tex
\section{Local Gaussian Process Hyperparameter Selection and Testing}

\begin{figure}[H]
    \centering
    \includegraphics[width = 14cm]{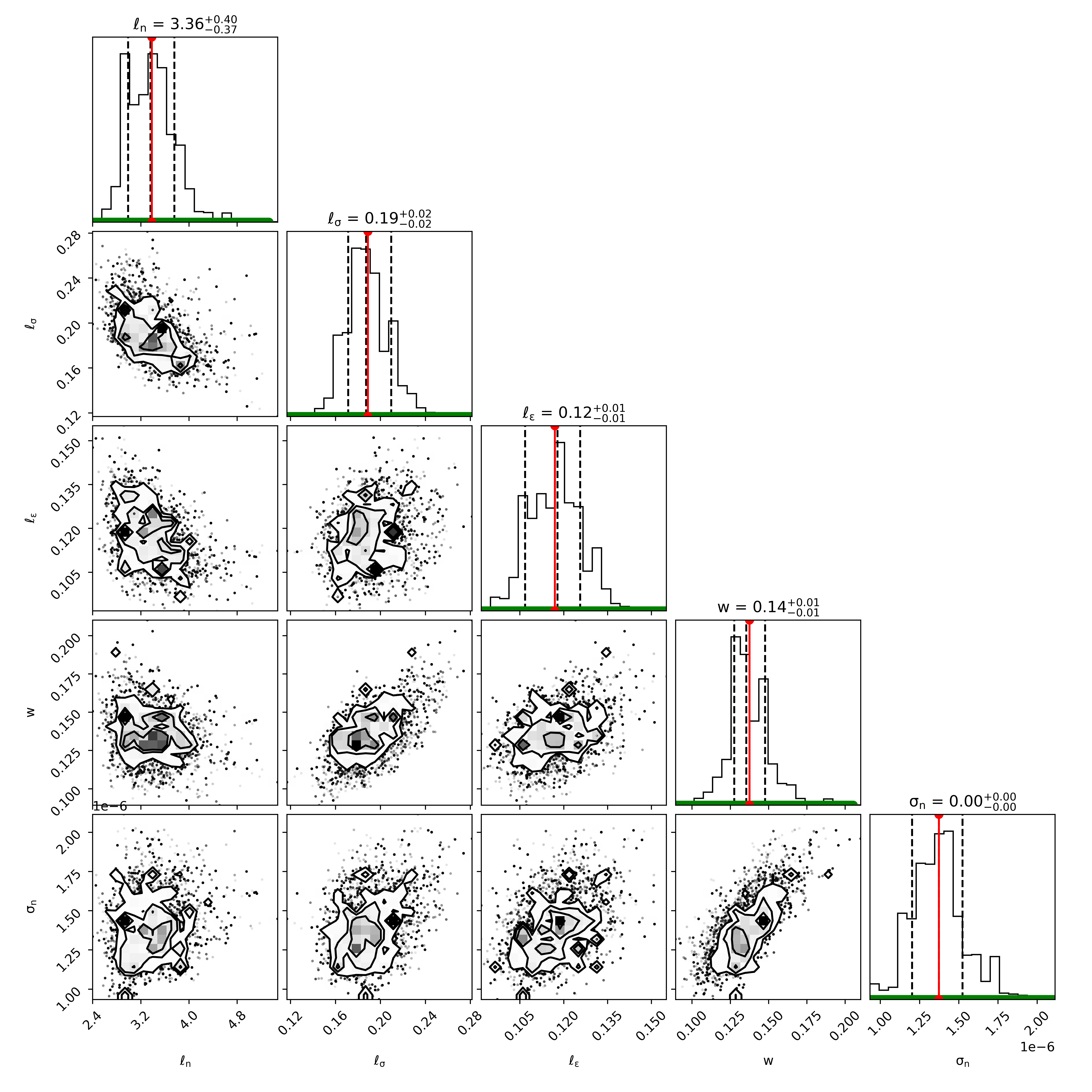}
    \caption{Leave-one-out marginal likelihood distributions estimated with MCMC sampling.}
    \label{fig:hyperparams}
\end{figure}

Local Gaussian process (LGP) surrogate model training for Mie fluid structure factors was performed using our recent method \cite{shanks_accelerated_2024}. The leave-one-out marginal likelihood approach from Sundararajan and coworkers \cite{sundararajan_predictive_2001} was used to estimate LGP hyperparameter distributions with Markov chain Monte Carlo (MCMC) sampling (see Figure \ref{fig:hyperparams}). MCMC samples were calculated using the emcee package \cite{foreman-mackey_emcee_2013} from 160 walkers with dynamic burn-in and sample time based on the autocorrelation convergence criterion (default stretch move with dynamic tuning). The \textit{maximum a posteriori} hyperparameter estimates (dashed red line) were selected as the surrogate model parameters.  

\begin{figure}[H]
    \centering
    \includegraphics[width = 14cm]{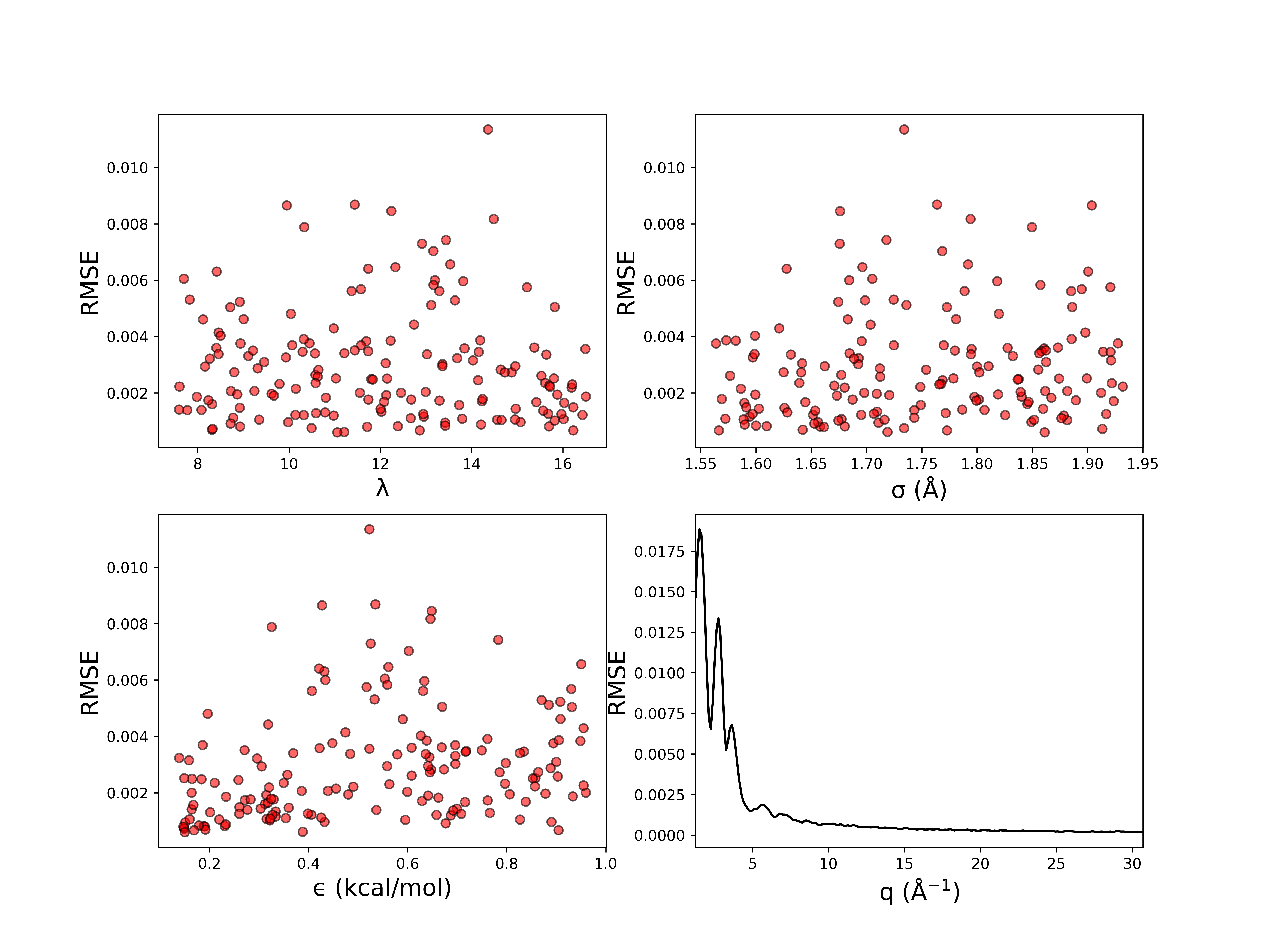}
    \caption{Root-mean-square error between LGP surrogate model prediction and training set points.}
    \label{fig:rmse}
\end{figure}

The LGP surrogate model was validated by generating a test set of 160 random samples within the convex hull of the training set defined according to Table \ref{tab:rangestest}. A reduced LGP test space was selected to avoid poor performance near the training boundaries. Boundaries for the LGP test set were determined by shifting the training set boundary by 12.5\% of the parameter training range. The root-mean-square-error (RMSE) was computed for each test condition (plots (a)-(c) in Figure \ref{fig:rmse}) and averaged over all structure factor $Q$ points (plot (d) in Figure \ref{fig:rmse}). The average RMSE over $Q$ was determined to be 0.0036. Surrogate model derivatives were computed only over this validated range.

\begin{table}[H]
\centering
\caption{LGP test set boundaries used for surrogate model validation.}
\begin{tabular}{| c | l | l |}
\hline
\textrm{Param.}&
\textrm{Min.}&
\textrm{Max.}
\\
\hline
$\lambda$  & 7.59 & 16.519 \\
$\sigma$   & 1.56 & 1.94 \\
$\epsilon$ & 0.14 & 0.96 \\
\hline
\end{tabular}
\label{tab:rangestest}
\end{table}

